\documentclass[11pt]{article}
\usepackage{moriond,epsfig}

\usepackage{multirow,color}

\def\be{\begin{equation}}
\def\ee{\end{equation}}
\def\bea{\begin{eqnarray}}
\def\eea{\end{eqnarray}}

\def\waa{{$W^\pm\gamma\gamma$}}
\def\wa{{$W^\pm\gamma$}}
\def\wz{{$W^\pm Z$}}

\def\sla#1{\ifmmode%
\setbox0=\hbox{$#1$}%
\setbox1=\hbox to\wd0{\hss$/$\hss}\else%
\setbox0=\hbox{#1}%
\setbox1=\hbox to\wd0{\hss/\hss}\fi%
#1\hskip-\wd0\box1 } 

\begin{document}
{\flushright{ 
    \begin{minipage}{20cm}
      DCPT/12/62, IPPP/12/31, FTUV-12-0511, KA-TP-18--2012, LPN12-051, SFB/CPP-12-26 \\
      \end{minipage}
  } 
} 
\vspace*{1cm}
\title{Precision Multiboson Phenomenology}
\vspace*{0.5cm}
\author{G.~Bozzi$^1$, F.~Campanario$^{2,~}$\footnote{Speaker, based on a
  talk given at the 47th Rencontres de Moriond on QCD and High Energy
  Interactions, March 10-17, 2012, La Thuile, Italy.}, C.~Englert$^3$, M.~Rauch$^2$, M.~Spannoswky$^3$, D.~Zeppenfeld$^2$}

\address{%
$^1$ {\it Dipartimento di Fisica, Universit\`a di Milano and INFN,
    Sezione di Milano\\ Via Celoria 16, I-20133 Milano, Italy}\\
$^2$ {\it Institute f\"ur Theoretische Physik, 
    Universit\"at Karlsruhe, KIT,\\ 76128 Karlsruhe, 
    Germany}\\
$^3$ {\it Institut for Particle Physics Phenomenology, Department of
  Physics, Durham University, Durham, United Kingdom }
}

\maketitle\abstracts{
We present recent results in precision multiboson (+jet) phenomenology at
the LHC. Results for diboson + jet, triboson, and also for \waa + jet will be
discussed focusing on the impact of the perturbative corrections on the
expected phenomenology.}
\section{Introduction}\label{Camp:sec:intro}
Processes with multiple electroweak bosons are important channels to test the
Standard Model (SM) at the LHC.  They are important backgrounds to SM and
also to beyond standard physics searches. As a signal, they allow us to obtain information on triple and quartic
couplings, and therefore, to quantify deviations from the SM
prediction through, e.g., anomalous couplings. 

To match the experimental accuracy, precise and reliable predictions
beyond the leading order~(LO) perturbative expansion are required not
only for cross sections but also for differential distributions. %
As
part of such a program we have, in the past, determined next-to-leading-order (NLO) QCD corrections for
the production cross sections of all combinations of three electroweak
bosons~\cite{Hankele:2007sb
,Bozzi:2011wwa%
}, to
$W\gamma$ +
jet~\cite{Campanario:2009um,Campanario:2010hv}
, $WZ$ + jet~\cite{Campanario:2010hp,Campanario:2010xn} and also to \waa
+ jet~\cite{Campanario:2011ud}, available in the VBFNLO package~\cite{Arnold:2008rz}. %
In all cases, leptonic
decays of the electroweak bosons were included in the calculations. For the
production of three weak bosons and also for \waa, these results were verified against
independent
calculations~\cite{Lazopoulos:2007ix,Binoth:2008kt,Baur:2010zf} which are available for on-shell bosons and neglecting Higgs boson
exchange. 

In these proceedings, we review results for
\waa, \waa~+ jet and \wa/\wz~+jet, including their leptonic decays and
full off-shell effects, in
Section~\ref{Camp:sec:waa},  \ref{Camp:sec:waaj} and \ref{Camp:sec:waj},
respectively, for the LHC at 14 TeV. We summarize in
Section~\ref{Camp:sec:con}. 
 
\section{\waa}\label{Camp:sec:waa}
Among the triple vector boson production channels, \waa~ has turned
out to be of particular interest.  \waa~production
is sensitive to the $W W \gamma$ and $WW\gamma\gamma$  vertices~\cite{quartic}. In addition,
a final state with two photons and missing transverse energy is relevant in a
variety of beyond the standard model scenarios~\cite{Campbell:2006wx}: in
gauge-mediated supersymmetry breaking, for instance, the neutralino is
often the next-to-lightest supersymmetric particle and decays into a
photon plus a gravitino, giving a signal of two photons and missing $E_T$. In
Ref.~\cite{CMS2009}, a study of the backgrounds for
supersymmetry motivated di-photon production searches has been performed,
pointing out the relevance of the \waa~production process as a SM background 
in case of electron misidentification. Another possible
application is an estimate of backgrounds when searching for $WH$ production,
followed by Higgs decay to two photons. 

We compute the NLO hadronic cross
section by straightforward application of the Catani-Seymour dipole
subtraction \cite{Catani:1996vz}. The loop contributions are evaluated using the
Passarino-Veltman scheme up to four-point functions \cite{Passarino:1978jh} and
the Denner-Dittmaier reduction \cite{Denner:2002ii} for five point integrals
and we perform various cross checks to validate our implementation, Refs.~\cite{Bozzi:2011wwa,Campanario:2011cs}.

\begin{figure}[h!]
\begin{center}
  \includegraphics[scale=0.7]{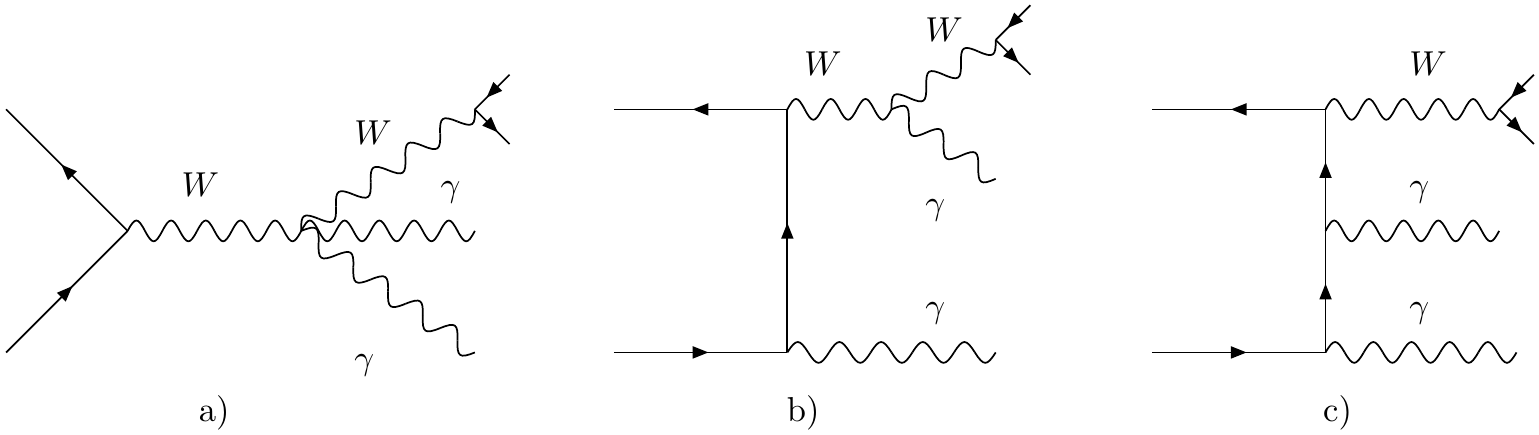}
\caption{\label{Camp:fig:LO} Example of the three topologies contributing to
  $pp \to l \nu \gamma\gamma$ + X }
\end{center}
\end{figure}

\begin{figure}[h!]
\begin{center}
  \includegraphics[scale=0.85]{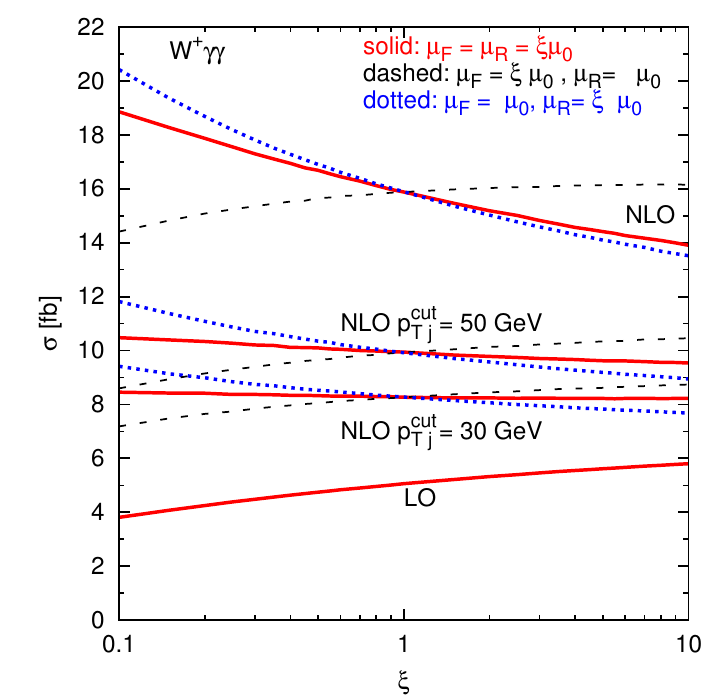}
\hspace*{1cm}
 \includegraphics[scale=0.85]{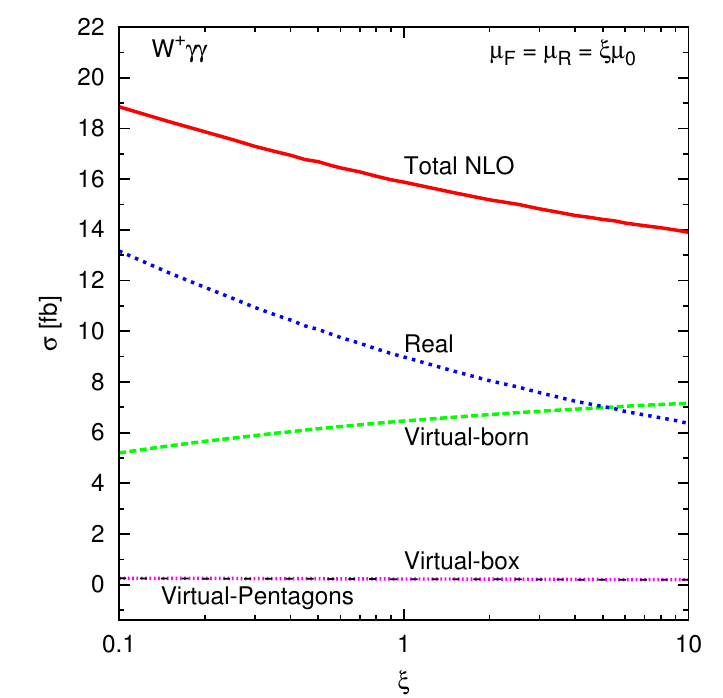}
\end{center}
\caption{\label{Camp:fig:waaT} {\it Left:} 
 Scale dependence of the total LHC cross section for 
      $p p \to \ell^+ \gamma \gamma +\sla{p}_T+X$ at
      at LO and NLO,  within the 
      cuts of Eqs.~(\ref{Camp:eq:cuts},\ref{Camp:eq:isol}).
      The factorization and renormalization scales are 
      varied in the range from $0.1 \cdot \mu_0$ to $10 \cdot
      \mu_0$. 
    {\it Right:}  Same as in the left panel but for the different NLO
      contributions at $\mu_F=\mu_R=\xi\mu_0$.}
\end{figure}

The NLO virtual corrections result from one-loop diagrams obtained by
attaching a gluon line to the quark-antiquark line in diagrams like the
ones depicted in
Fig.~\ref{Camp:fig:LO}. We combine the virtual corrections into three
different groups, which  include all loop diagrams derived from a given
Born level configuration. This leaves us with three
universal building blocks, namely factorizable corrections (Virtual-born) and
corrections to two~(Virtual-box) or
three ~(Virtual-Pentagons) vector bosons attached
to the quark line. For our numerical results, we use the CT10 parton
distribution set \cite{Lai:2010vv} with $\alpha_s(m_Z)=0.118$ at NLO,
and the CTEQ6L1 set~\cite{Pumplin:2002vw} with $\alpha_s(m_Z)=0.130$
at LO.  We impose a set of
minimal cuts on leptons, photons and jets, namely,

\begin{equation}
  p_{T{\ell(\gamma)}} > 20 \ \mathrm{GeV} \qquad
  |y_{\ell(\gamma)}| < 2.5 \qquad
  R_{\gamma\gamma} > 0.4  \qquad
  R_{\ell\gamma} > 0.4  \qquad
  R_{j \ell} > 0.4  \qquad
  R_{ j \gamma} > 0.7 
  \label{Camp:eq:cuts}
\end{equation}
as well as an isolation criteria \`a la Frixione~\cite{Frixione:1998jh} for the photons,
 \be
\Sigma_i \, E_{T_i} \, \theta (\delta - R_{i\gamma}) \, \leq \,
p_{T\gamma} \, \frac{1-\cos\delta}{1-\cos\delta_0} \,\,\,\,\,\,\,\,\,\,
(\mathrm{for\,all} \,\,\,\,\,\delta\leq\delta_0), \label{Camp:eq:isol}
\ee
where $\delta_0$ is a fixed separation which is set to 0.7.
We consider $W^\pm$ decays to 
the first two lepton generations, i.e., $W\rightarrow e\nu_e(+\gamma),\mu\nu_\mu(+\gamma)$ and these
contributions have been summed in Fig.~\ref{Camp:fig:waaT}, where  we show numerical results for
$W^+\gamma\gamma$ production within the cuts of Eqs.~(\ref{Camp:eq:cuts}, \ref{Camp:eq:isol}).
On the left panel, we show the overall scale variation of
our numerical predictions at LO and NLO: the NLO K-factor is large both
in absolute value  ($\sim$ 3) and compared to the LO scale variation. The NLO scale
uncertainty is about 10$\%$ when varying the factorization and the
renormalization scale $\mu=\mu_F=\mu_R$ up and down by a factor 2 around
the reference scale $\mu_0=m_{W\gamma\gamma}$ and is mainly driven by the dependence on
$\mu_R$. 
The large size of the NLO corrections partially originates from new gluon
induced channels entering first at NLO, $gq\to$ \waa~$q$, which are
$\alpha_s$ suppressed, but enhanced by the large gluon pdfs at the LHC. 
Since these 1-jet contributions to the ${\cal O} (\alpha_s)$ cross section
are only determined at LO, and are unbalanced against the virtual part, 
their scale variation is large. In fact, most
of the scale variation of the total NLO result is
accounted for by the real emission contributions, defined here as the
real emission cross section minus the Catani-Seymour subtraction terms
plus the finite collinear terms. This is more visible in the right
panels, where we show the scale dependence and compare the
size of the different parts of the NLO calculation.  As for the relative size of the NLO
terms, the real emission contributions dominate and are even larger than
the LO terms plus virtual terms proportional to the Born
amplitude. Non-trivial virtual contributions, namely the interference of
the Born amplitude with virtual-box and virtual-pentagon contributions,
represent less than 1\% of the total result and their scale 
dependence is basically flat. In the left panels,
we also show results for additional jet veto cuts, requiring $p_{T j}<50$~GeV
or $p_{T j}<30$~GeV.
While it is evident that the renormalization scale variation is highly reduced by
a jet veto, this reduction should not be interpreted as a smaller
uncertainty of the vetoed cross section: a similar effect in \wa$j$ and
$WZj$ and \waa$j$ production could be traced to cancellations between different regions of
phase space and, thus, the small variation is
cut-dependent~\cite{Campanario:2010hp} as shown in the following sections.

 Among  the triple vector boson production channels,  \waa~production is the one
 with  the largest K-factor for the integrated cross section. In
 Ref.\cite{Bozzi:2011wwa} (see also \cite{Baur:2010zf}), it was shown that
 this is due to cancellations at LO driven by a radiation
 zero~\cite{Brown:1982xx}. The radiation zero at NLO is
 obscured, similar to $W^\pm \gamma $ production~\cite{Baur:1993ir}, by
 additional real QCD radiation, \waa~+jet, as
part of the NLO contributions. An
additional jet veto-cut might help in the detection of the radiation
zero, while reducing also the scale uncertainties
for the relevant distributions. However, this procedure raises the
question of the reliability of the predictions due to the aforementioned
problem with the exclusive vetoed samples. Furthermore, the remaining
scale uncertainties at NLO QCD are due to unbalanced
gluon-induced real radiation 
computed at LO, e.g., $gq\to$ \waa~$q$. To realistically asses the
uncertainties, also concerning anomalous coupling searches, and as an important step towards a NNLO QCD calculation of
\waa~, we have calculated \waa~+jet at NLO QCD. 


\section{\waa + jet}\label{Camp:sec:waaj}

This is the first
calculation falling in the category of $VVV+j$, which includes the
evaluation of the complex hexagon virtual amplitudes, which poses a
challenge not only at the level of the analytical calculation, but also
concerning the CPU time required to perform a full $2\to 4$ process at
NLO QCD. 

For the virtual contributions we use the routines computed
in Ref.~\cite{Campanario:2011cs}. 
At the numerical evaluation
level, we split the virtual contributions
into fermionic loops~(Virtual-fermionbox) and bosonic contributions with one~(Virtual-box),
two~(Virtual-pentagons) and three~(Virtual-hexagons) electroweak vector bosons attached to the quark line.
This procedure allows us to drastically reduce the time spent in
evaluating the part containing hexagon diagrams as explained in
Refs.~\cite{Campanario:2011ud,Campanario:2011cs}. The numerical
stability of the hexagons' contributions is discussed in detail in
Ref.~\cite{Campanario:2011cs}.

\begin{figure*}[h!]
  \includegraphics[width=0.5\columnwidth]{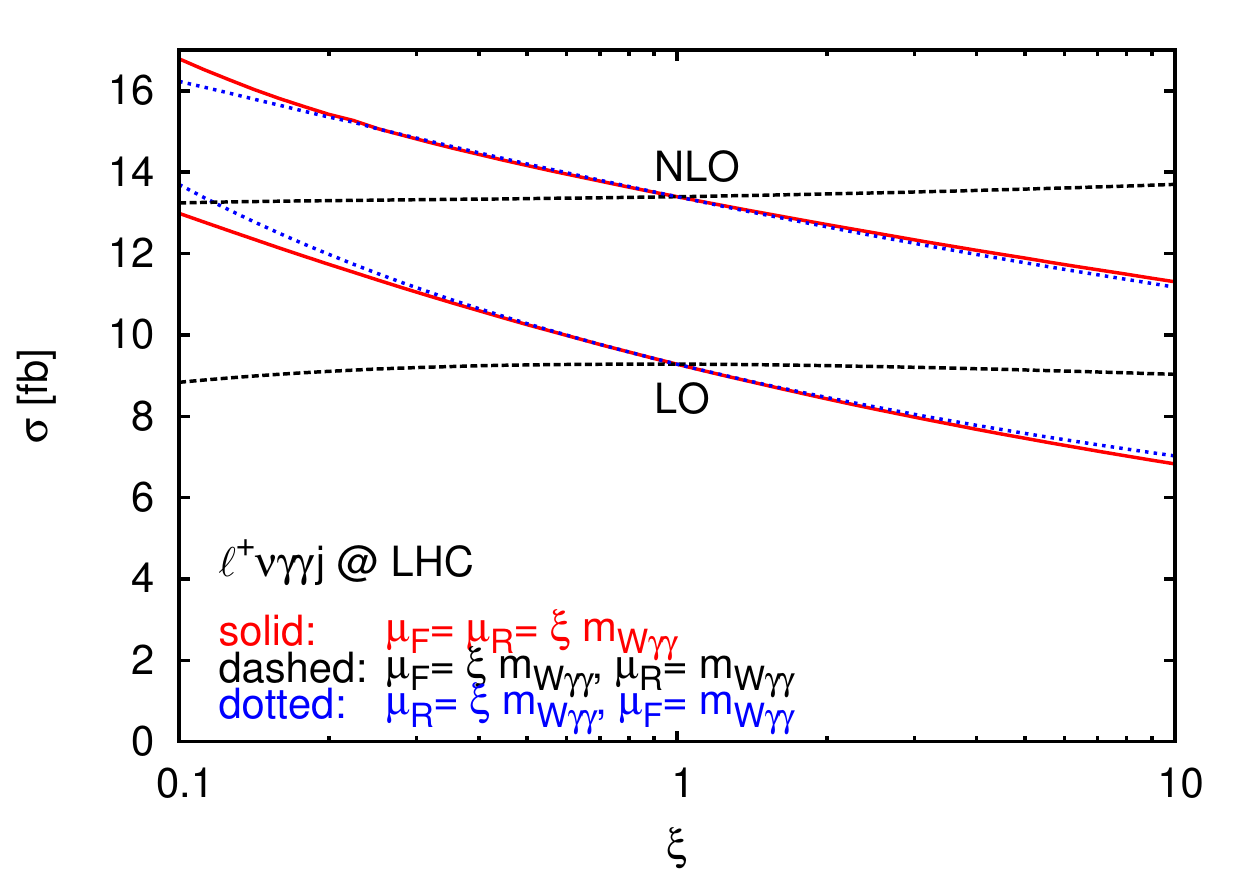}
  \hfill
  \includegraphics[width=0.5\columnwidth]{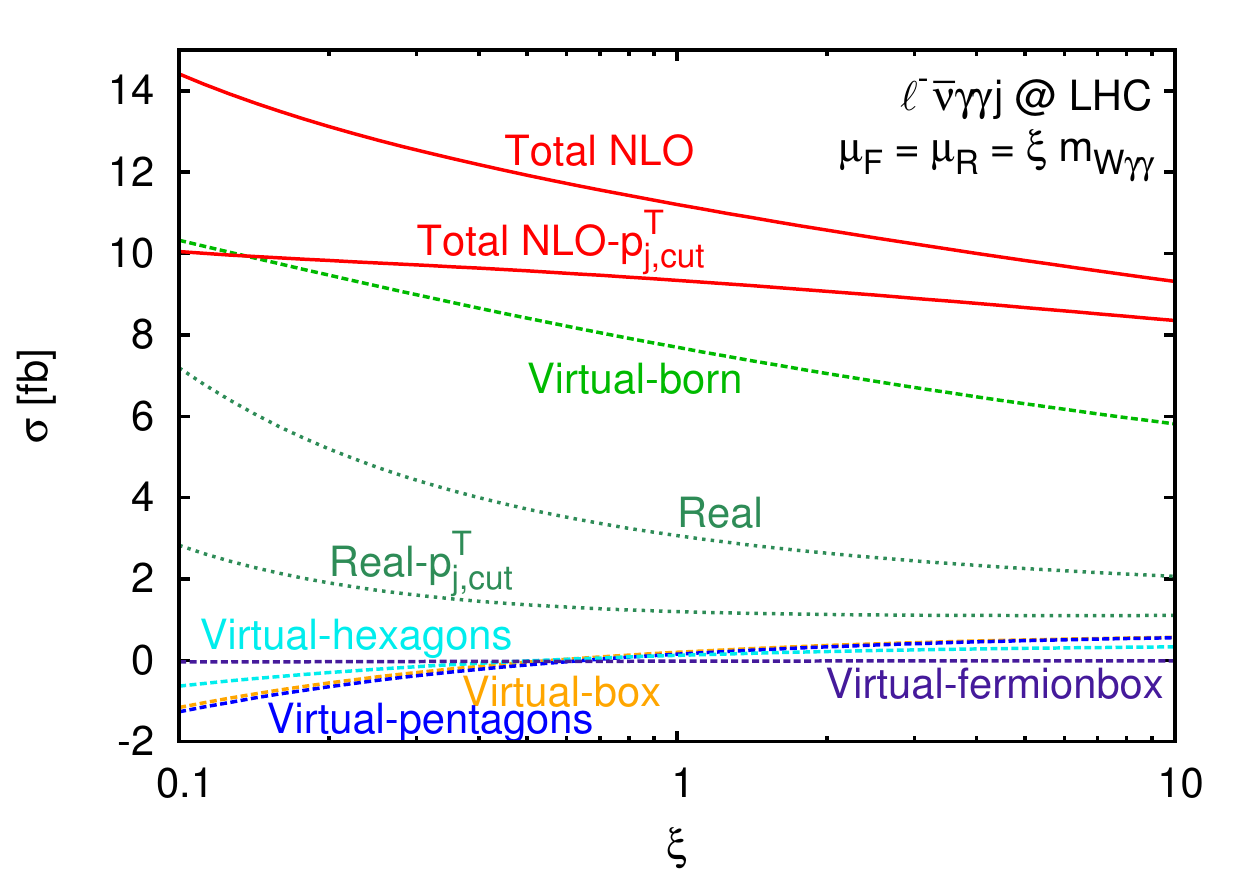}
  \caption{\label{Camp:fig:waaj:scale} Scale variation of the $\ell^\pm
      \nu \gamma\gamma$+jet production cross sections at the LHC
      ($\ell$ = $e$, $\mu$). The cuts are described in the
      text and we choose $\mu_R=\mu_F=m_{W\gamma\gamma}$ as central
      dynamical reference scale. The right panel shows the individual
      contributions to the NLO cross section according to our
      classification of topologies. We also show results where we have
      applied a veto on events with two identified jets having both
      $p_{\rm{T}}^j> 50$ GeV} 
\end{figure*}

\begin{figure}[!h]
\begin{center}
 \includegraphics[height=0.5\columnwidth]{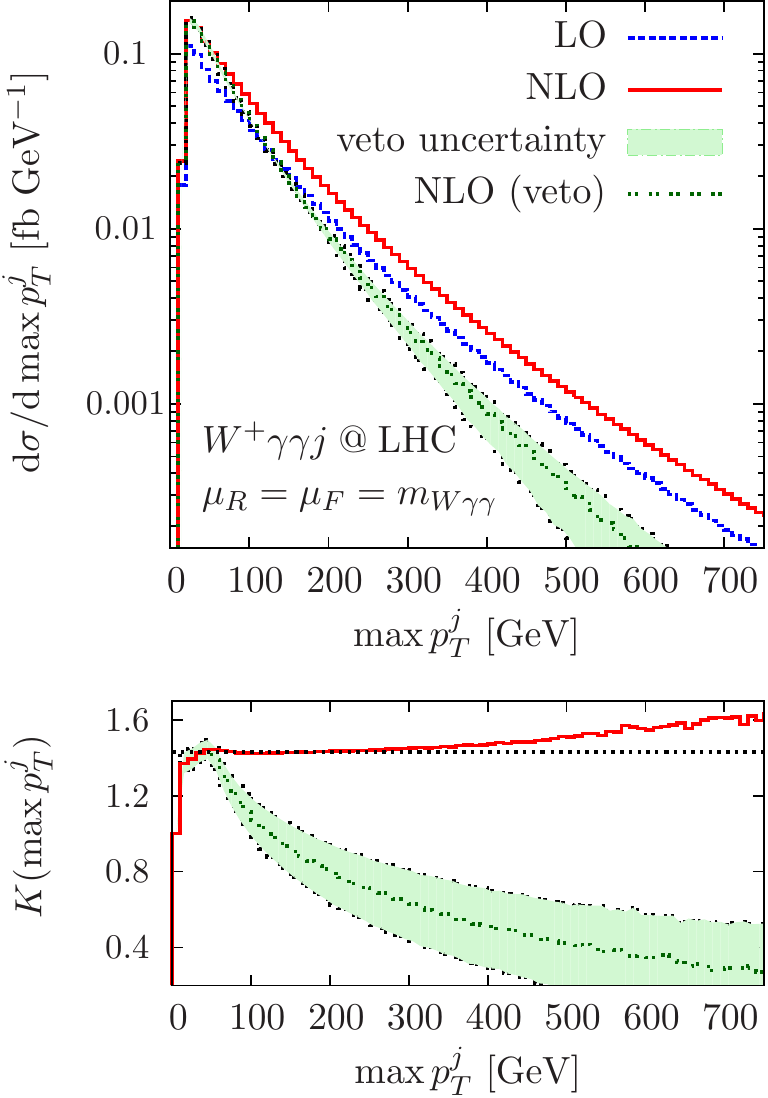}
\hspace*{1cm}
 \includegraphics[height=0.5\columnwidth]{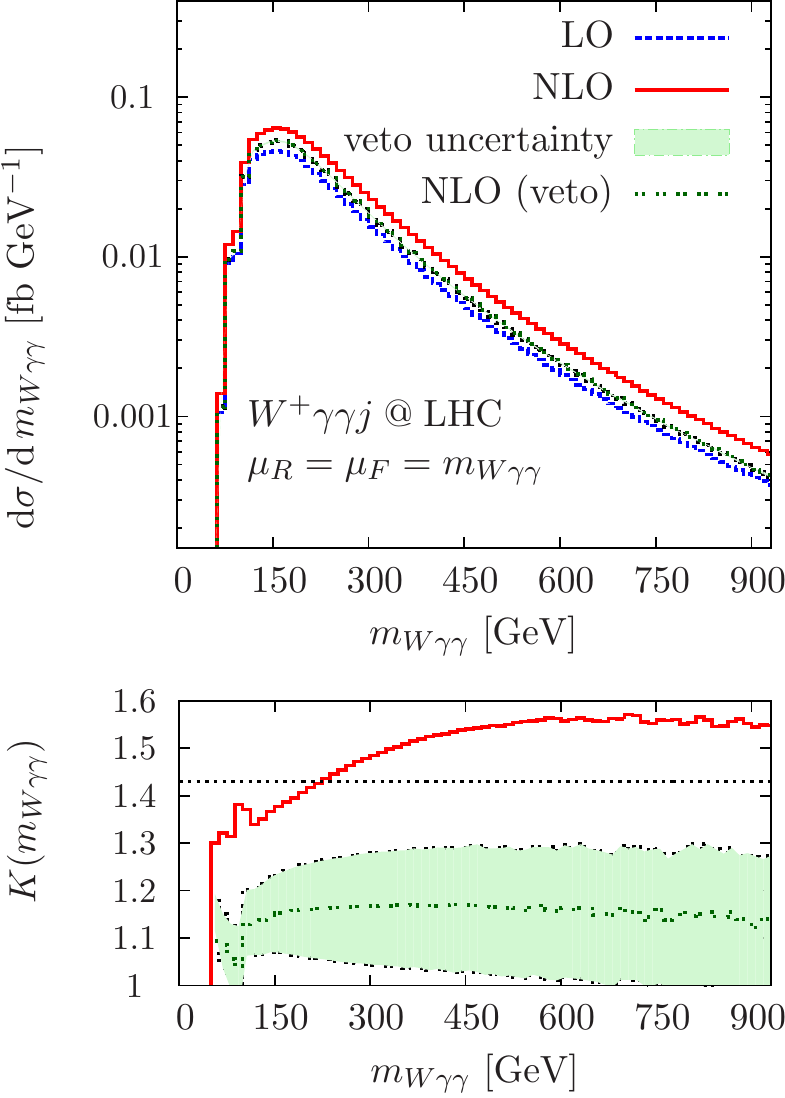}
\end{center}
  \caption{\label{Camp:fig:waaj:diff} Differential $\max p_T^j$ and
      $m_{W\gamma\gamma}$ distribution for inclusive and exclusive
      $l^+\bar \nu \gamma\gamma$+jet production.}
    \vspace{0.1cm}
\end{figure}

We use the same input parameters as for \waa~production and apply the cuts of
Eqs.~(\ref{Camp:eq:cuts},\ref{Camp:eq:isol}). 
Further details on the parameter choices can be found in 
Ref.~\cite{Campanario:2011ud}.  Again, we consider $W^\pm$ decays to 
the first two lepton generations, i.e, $W\rightarrow
e\nu_e(+\gamma),\mu\nu_\mu(+\gamma)$ and these contributions have been summed in Fig~\ref{Camp:fig:waaj:scale} and~\ref{Camp:fig:waaj:diff}.

We compute total $K$ factors of 1.43 (1.48) for
$W^+\gamma\gamma$+jet ($W^-\gamma\gamma$+jet) production at the
LHC, values which are quite typical for multiboson+jet production as
found in Refs.~\cite{Campanario:2010hv,Campanario:2010xn,dibosjet} and
partially originated by new e.g, $gg$ and $qq$ induced channels.  This moderate K-factor~(as compared
  to corrections of  $\sim 300\%$ for \waa~production) indicates,
   as expected, that the \waa~+ jet production channel is not affected by
   radiation zero cancellations. 

The scale dependences of the $W^+\gamma\gamma j$ and $W^-\gamma\gamma j$
production cross sections turn out to be modest: when comparing 
$\mu_R=\mu_F=\xi m_{W\gamma\gamma}$ for $\xi=0.5$ and $\xi=2$, we find differences of
$10.8\%$ ($12.0\%$), respectively, see Fig.~\ref{Camp:fig:waaj:scale}.

The phase space dependence of the QCD corrections is non-trivial and
sizable (we again choose $\mu_R=\mu_F=m_{W\gamma\gamma}$). Vetoed
real-emission distributions are plagued with large
uncertainties (Fig.\ref{Camp:fig:waaj:diff}, left) --- a
characteristic trait well-known from $VV$+jet phenomenology
\cite{Campanario:2010hp,dibosjet}. Additional parton emission modifies
the transverse momentum and invariant mass spectra in particular.
The leading jet becomes
slightly harder at NLO as can be inferred from the differential $K$
factor in the bottom panel of Fig.~\ref{Camp:fig:waaj:diff}. When comparing
precisely measured distributions in this channel against LO Monte Carlo
predictions, the not-included QCD corrections could be misinterpreted
for anomalous electroweak trilinear or quartic couplings
\cite{Campanario:2010hv,Campanario:2010xn,Baur:1993ir} arising from
new interactions beyond the SM.

\section{\wa/\wz~+ jet}\label{Camp:sec:waj}
\begin{figure}[h!]
\begin{center}
  \includegraphics[scale=0.65]{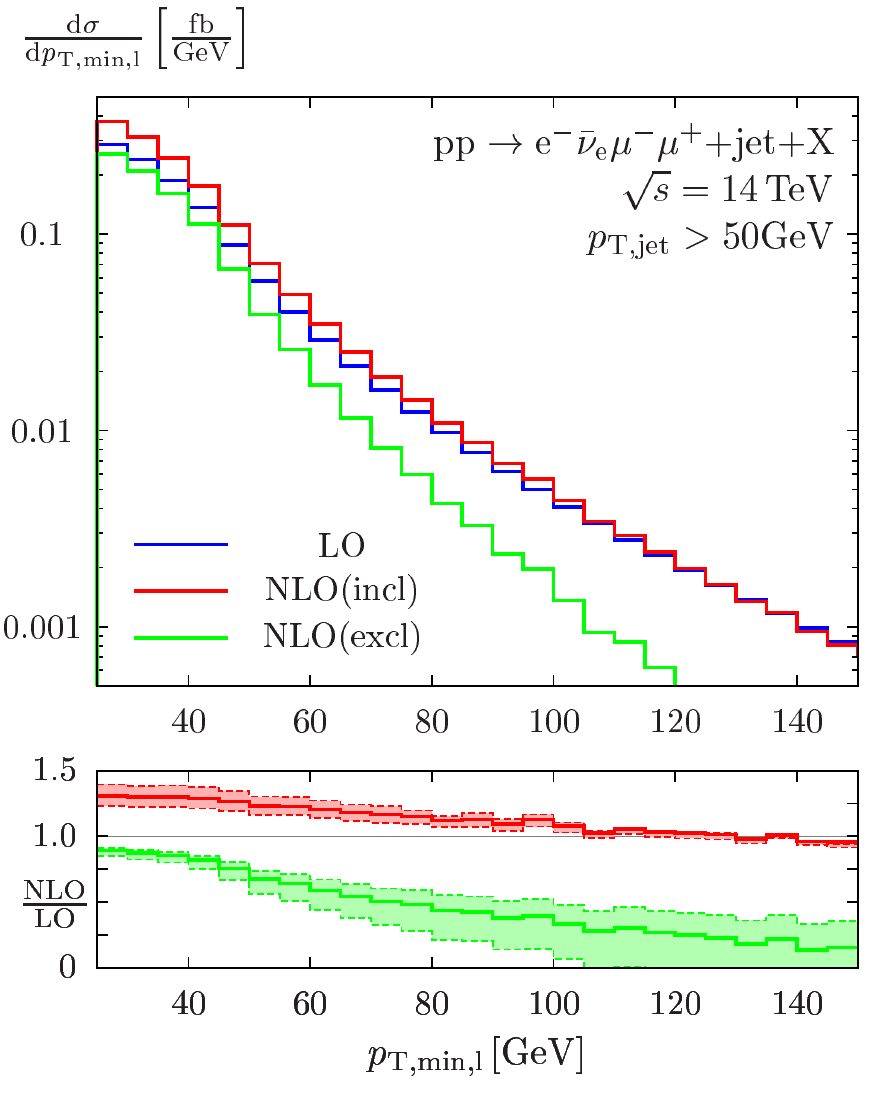}
\hspace*{1cm}
 \includegraphics[scale=0.75]{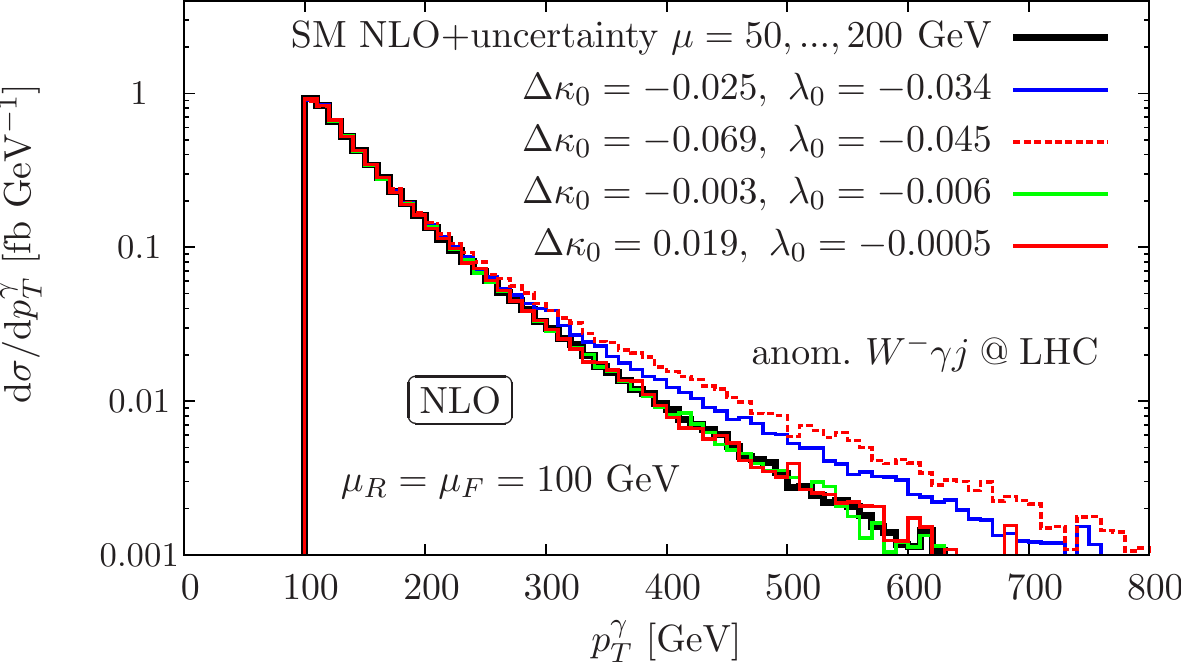}
\end{center}
\caption{\label{Camp:fig:wa} 
{\it Left:} Differential distribution for $p_{T,min,l}$ for inclusive and exclusive
      $W^-Z$+jet production. {\it Right:} Sensitivity to anomalous
      couplings for $l^-\nu \gamma$ + jet in the $p_{T}^\gamma$ distribution.}
\end{figure}

NLO corrections to $pp \to$\wa/\wz~+jet cross section have been computed in Refs.~\cite{Campanario:2009um,Campanario:2010hp}, and including
anomalous couplings in Refs.~\cite{Campanario:2010hv,Campanario:2010xn}.
All off-shell effects were included. Similar observations as in
\waa~+ jet were found. When varying the factorization and
renormalization scale by a factor 2 around fixed values of $\mu_0$=100
GeV, one finds modest scale variations. Vetoed samples pick up large 
uncertainties~(Fig.\ref{Camp:fig:wa}, left). K-factors are around 1.4 at the
LHC and they vary over phase space. Two examples are shown for these
processes in Fig.~\ref{Camp:fig:wa}, including the sensitivity to anomalous
couplings in the $p_T^\gamma$ differential distributions for \wa~+jet
production for different choices of anomalous parameters~($\lambda_0,k_0$).

\section{Summary}\label{Camp:sec:con}

The QCD corrections for vector boson
production in the diboson + jet, triboson and triboson + jet channels 
are large and exceed the expectations driven by LO scale
uncertainties. 
Total K-factors up to 3 for \waa~have been
reported. The size of the QCD corrections for \wa/\wz~+ jet and \waa~+ jet
production is around the 40$\%$ level. Corrections can be larger for
differential distributions and therefore have to be considered for a
precise comparison of data to SM predictions for all these
processes. Finally, we have shown that the diboson + jet production channels
are sensitive to anomalous coupling searches through differential distributions.

\end{document}